\documentclass[12pt,preprint]{aastex}
\usepackage{graphicx}

\newcommand{\ergsacm}{erg~s$^{-1}$~cm$^{-2}$ ~\AA$^{-1}$}

\def\etal{et\thinspace al.\ }                               

\shortauthors{Gonz\'alez Delgado et al.}
\shorttitle{Stellar population of LLAGN}

\begin{document}


\title{Stellar Population in LLAGN.II: STIS observations
\footnote{Based on observations with the NASA/ESA {\it Hubble Space Telescope}, 
obtained at the Space Telescope Science Institute, which is operated 
by the Association of universities for Research in Astronomy, Inc., 
under NASA contract NAS 5-26555. 
Based on observations made with the Nordic Optical Telescope, 
operated on the island of La Palma jointly by Denmark, Finland, 
Iceland, Norway, and Sweden, in the Spanish Observatorio del 
Roque de los Muchachos of the Instituto de Astrof\'\i sica de Canarias.}
}

\author{Rosa M. Gonz\'alez Delgado\altaffilmark{1}, 
Roberto Cid Fernandes\altaffilmark{2},  
Enrique P\'erez\altaffilmark{1}, Lucimara P. Martins\altaffilmark{3,7}, 
Thaisa Storchi-Bergmann\altaffilmark{4}, Henrique Schmitt\altaffilmark{5,8},
Timothy Heckman\altaffilmark{6,9}, \& Claus Leitherer\altaffilmark{3}}

\affil{(1) \em Instituto de Astrof\'{\i}sica de Andalucia (CSIC), P.O. Box 3004, 18080 Granada, Spain (rosa@iaa.es; eperez@iaa.es)}
\affil{(2) \em Depto. de F\'\i sica-CFM, Universidade Federal de Santa 
Catarina, C.P. 476, 88040-900, Florian\'opolis, 
SC, Brazil (cid@astro.ufsc.br)}
\affil{(3) \em Space Telescope Institute, 3700 San Martin Drive, Baltimore, MD 21218, USA (martins@stsci.edu; leitherer@stsci.edu)}
\affil{(4) \em National Radio Astronomy Observatory, PO Box 0, Socorro, NM 87801 (hschmitt@nrao.edu)}
\affil{(5) \em Instituto de F\'\i sica, Universidad Federal do Rio Grande do Sul, C.P. 15001, 91501-970, 
Poto Alegre, RS, Brazil (thaisa@if.ufrgs.br)}
\affil{(6) \em Department of Physics \& Astronomy, JHU, Baltimore, MD 21218 (heckman@pha.jhu.edu)}
\affil{(7) \em Intituto de Astronom\'\i a, Geof\'\i sica e Ciencias Atmosf\'ericas, 05508-900  Sao Paulo, Brazil}

\altaffiltext{8}{Jansky Fellow} 
\altaffiltext{9}{Also Adjunct Astronomer at STScI} 
\begin{abstract}

We present a study of the stellar population in Low Luminosity Active
Galactic Nuclei (LLAGN). Our goal is to search for spectroscopic
signatures of young and intermediate age stars, and to investigate
their relationship with the ionization mechanism in LLAGN. The method
used is based on the stellar population synthesis of the optical
continuum of the innermost (20-100 pc) regions in these galaxies. For
this purpose, we have collected high spatial resolution optical
(2900-5700 \AA) STIS spectra of 28 nearby LLAGN that are available in
the {\it Hubble Space Telescope} archive.  The analysis of these data
is compared with a similar analysis also presented here for 51
ground-based spectra of LLAGN. Our main findings are: (1) No features
due to Wolf-Rayet stars were convincingly detected in the STIS
spectra. (2) Young stars contribute very little to the optical
continuum in the ground-based aperture. However, the fraction of light
provided by these stars is higher than 10\% in most of the weak-[OI]
([OI]/H$\alpha\leq$ 0.25) LLAGN STIS spectra. (3) Intermediate age
stars contribute significantly to the optical continuum of these
nuclei. This population is more frequent in objects with weak than
with strong [OI].  Weak-[OI] LLAGN that have young stars stand out for
their intermediate age population.  (4) Most of the strong-[OI] LLAGN
have predominantly old stellar population. A few of these objects also
show a feature-less continuum that contributes significantly to the
optical continuum. These results suggest that young and intermediate
age stars do not play a significant role in the ionization of LLAGN
with strong [OI]. However, the ionization in weak-[OI] LLAGN with
young and/or intermediate age population could be due to stellar
processes. A comparison of the properties of these objects with
Seyfert 2 galaxies that harbor a nuclear starburst, suggests that
weak-[OI] LLAGN are the lower luminosity counterparts of the Seyfert 2
composite nuclei.
 
\end{abstract}



\section{Introduction}

Low-Luminosity Active Galactic Nuclei (LLAGN) constitute a sizeable
fraction of the nearby AGN population.  These include low-luminosity
Seyferts, low-ionization nuclear emission-line regions (LINERs), and
transition-type objects (TOs) whose properties are in between
classical LINERs and HII nuclei. LLAGN comprise about 1/3 of all
bright galaxies (B$_T$ $\leq$ 12.5) and are the most common type of
AGNs (Ho, Filippenko \& Sargent 1997a; hereafter HFS97).

TOs could constitute a rather mixed phenomena as suggested by the
several excitation mechanisms that have been proposed to explain the
origin of their energy source.  Among these mechanisms are shocks,
photoionisation by a non-stellar UV/X-ray continuum (AGN), and
photoionisation by hot stars (see e.g.\ review by Fillipenko 1996).
The possibility that some LINERs (as well as Seyferts) might be
photoionised by hot stars has been suggested by several authors (e.g.\
Stasi\'nska 1984, Filippenko \& Terlevich 1992, Shields 1992). These
hot stars could be the product of the evolution of massive stars (the
so-called {\it warmers} proposed by Terlevich \& Melnick 1985) or of
intermediate mass stars (the post-AGB stars investigated by Binette et
al. 1994 by Taniguchi, Shioya, \& Murayama 2000).

The massive star scenario has been reexamined by Barth \& Shields
(2000), using detailed starburst plus photoionisation modeling with
evolutionary stellar synthesis models. These authors showed that the
TOs properties can be reproduced by short stellar bursts with ages of
$\sim$ 3-6 Myr, when the ionizing continuum is dominated by emission
from Woft-Rayet (WR) stars. More recent photoinization models with
updated Starburst99 (Smith, Norris, \& Crowther 2002), which include new blanketed WR
and O atmospheres, are not able to reproduce the typical TOs emission
line ratios (Gonz\'alez Delgado et al 2003).

There is further evidence in favor of the starburst scenario, coming
from the detection of stellar wind lines in the ultraviolet spectra of
some weak-[OI] LINERs and TOs (Maoz et al 1998; Colina et al 2002). In
fact, similar spectroscopic features detected in Seyfert 2 galaxies
are interpreted as due to a few Myr old nuclear starbursts (Heckman et
al 1997; Gonz\'alez Delgado et al 1998).

About 20$\%$ of LLAGN in the catalogue of HFS97 required a broad component to fit the H$\alpha$ emission
(type 1 LINERs; Ho et al 1997b). Some of these objects show
double-peaked H lines (Storchi-Bergmann et al 1997; Shields et al
2000; Ho et al 2000). In addition, recent X-ray observations of LLAGN
confirm that some objects are at the low-luminosity end of the AGN
phenomenon, and that they are powered by an accreting black-hole
(BH) (Terasima, Ho, Ptak \&  2000; Ho et al 2001).

Furthermore, recent high spatial resolution UV (Colina et al 2002) and
Chandra (Jim\'enez-Bail\'on et al 2003) observations of the LLAGN NGC
4303 shows that a super-stellar cluster (SSC) and a BH accreting with
low radiative efficiency coexist within the inner few pcs from the
nucleus.

Considering the diversity of excitation mechanisms that can explain
the emission line spectrum of LLAGN, we have started a project to
examine the central stellar population in these objects, with the aim
of finding clues about their physical origin and energy source. With
this goal, we have carried out optical (3500--5500 \AA) spectroscopic
observations of LLAGN to: (1) search for the presence of the broad WR
bump at the blue optical range (a blend of broad He\,{\sc ii }$\lambda$4686, N\,{\sc iii}
$\lambda$4640 and C\,{\sc iv} $\lambda$4650); (2) detect the
absorption lines of HeI and the high order HI Balmer series (HOBLs);
(3) characterize their stellar populations. The WR bump probes the
presence of very young stars (few Myr old), while HeI and HOBLs probe
the young (10--50 Myr) and intermediate (100--1000 Myr) age stars
(Gonz\'alez Delgado, Leitherer, \& Heckman 1999). In addition, this spectral range
contains many metallic stellar lines typical of old and intermediate
age populations.

In a companion paper (Cid Fernandes \etal 2003a, hereafter Paper I),
we have presented the data and an empirical correlation between the
stellar lines and the emission lines in LLAGN, in particular the
[OI]/H$\alpha$ line ratio. Our main findings from Paper I are: (1) No
features due to WR stars are convincingly detected, implying that
massive stars contribute very little to the optical light. This
happens even in the few cases where young stars are known to dominate
the UV emission. (2) HOBLs, on the other hand, are detected in
40\% of the sample. These
lines are absent in most strong-[OI] LLAGN, but they are detected in
about 50$\%$ of the weak-[OI] LLAGN, defined as TOs and LINERs with
[OI]/H$\alpha \leq 0.25$. In fact, $\sim 90\%$ of nuclei exihibiting
HOBLs are weak-[OI] emitters.

The analysis in Paper I was carried out entirely in empirical terms,
investigating connections between observed quantities. In this paper
we present a stellar population synthesis analysis of the nuclear
spectra of these objects.  Because our results may be affected by the
spatial resolution of the ground-based observations, we also present
here the analysis of 28 LLAGN observed at optical wavelengths with the
Space Telescope Imaging Spectrograph (STIS) on board the Hubble Space
Telescope (HST). The high spatial resolution provided by HST+STIS may
be crucial to spatially-isolate the light from a central compact
source or different circumnuclear star forming knots from the
underlying light emitted by the old stars in the inner bulge of these
galaxies.

Another major goal of this study is to investigate the low luminosity
end of the ``Starburst-AGN connection''. Powerful circumnuclear
starbursts are present in 30--50\% of type 2 Seyferts (Cid Fernandes
\etal 2001a and references therein). The statistics are nor clear for
type 1 Seyferts due to the difficulty in spotting circumnuclear
starbursts against the bright nucleus (which is conveniently obscured
in Seyfert 2s), but if unification is correct the fraction of
starburst+AGN composites should be similar in Seyfert 1s and 2s. If
LINERs are just ``mini-Seyferts'', one would naively expect to find
circumnuclear starbursts in about 30--50\% of them. Different
incidence rates could be due to evolutionary effects, which, if
present, should also be detected in a comparative stellar population
analysis of Seyfert 2s and LINERs.

This paper is organized as follows: In section 2 we describe the
sample and the STIS observations. Section 3 presents the STIS spectra
and measurements of their properties. An empirical population
synthesis (EPS) analysis of the whole sample (ground-based + STIS) is
presented in section 4. The results of this analysis are discussed in
section 5, where we investigate the connection between the inferred
stellar populations and emission line properties, compare our results
with those obtained for Seyfert 2s and Starburst galaxies, and
speculate on possible evolutionary scenarios.  Finally, section 6 
summarizes our conclusions.
 
\section{Observations and data reduction}
 
\subsection{Galaxy Sample}

Our sample of LLAGN was drawn entirely from the HFS97 catalogue because it is the most complete and
homogeneous survey of LINERs and TOs available for the local
universe. In addition to the galaxies presented in Paper I, we have
selected LLAGN observed with STIS in the 2900--5700 \AA\ spectral
range. Spectra of 32 LLAGN are available in the HST archive, of which
28 galaxies (17 TOs and 11 LINERs according to the classification of
HFS) are suitable for a stellar population analysis. Eight of these
TOs have also been observed from the ground by us.  These data are
complemented with STIS spectra of four galaxies classified by HFS
as HII nuclei, plus NGC 1023, a non-active galaxy. Most of these
observations are from proposals number 8607 (P.I.  L.C. Ho) and 7361
(P.I. H-W. Rix). These projects include 24 nearby early type (S0--Sb)
galaxies and 15 TOs from the HFS97 catalogue which are closer than 17
Mpc and have emission lines with fluxes larger than 10$^{-15}$
erg~s$^{-1}$~cm$^{-2}$ ~\AA$^{-1}$ in a 2$\times$4 arcsec nuclear
aperture.  Observations of NGC 1023, NGC 3507, NGC 3998, and NGC 4261
are from proposals 7566 (P.I. R. Green), 7357 (P.I. L.C. Ho), 8839
(P.I. L. Dressel) and 8236 (P.I. S. Baum), respectively. Properties of
these objects are listed in Table 1.

This collection plus data from Paper I increases the sample analized
here to 33 TOs and 40 LINERs, which represents 49\% of the TOs and
42\% of the LINERs in the HFS97 sample. If we adopt a slightly different
classification criterium for LLAGN, as we did in Paper I, placing the
dividing line between the two subtypes at [OI]/H$\alpha$=0.25, our
sample comprises 24 strong-[OI] and 47 weak-[OI] LLAGN, corresponding
to 43\% and 44\% of these types of galaxies in the HFS97 sample.
Morphological type and distance distributions are presented and
compared with the full HFS97 sample in Figure 1. Our sample has the same
median morphological type that the HFS97 sample; S0 and Sab for strong
and weak-[OI] LLAGN, respectively.  With respect to the distance, the
median in our sample is 17 Mpc and 23 Mpc for weak-[OI] and
strong-[OI] LLAGN, respectively. The corresponding values in the HFS
sample are 20 Mpc and 22 Mpc.

\subsection{Observations}

The observations were obtained with the STIS/CCD detector with a
52$\times$0.2 arcsec slit (except for NGC 3507, which was observed
with a 0.5 arcsec slit) and the G430L grating. The spectra cover the
wavelength range 2900--5700 \AA\ with a dispersion of 2.7 \AA/pixel,
giving a minimum FWHM spectral resolution of $\sim$ 4 \AA\ for point
sources and up to $\sim$11 \AA\ for extended sources. After an initial
acquisition exposure of a few seconds through the optical long-pass
filter, the slit was placed across each nucleus at a randomly oriented
position angle. The CCD spatial scale is 0.05 arcsec/pixel.
Observations from proposals 8607 and 8236 were binned every two
pixels, yielding a spatial sampling of 0.1 arcsec.

The data were calibrated with the standard STScI pipeline, meaning
that the spectra were bias and dark subtracted, flat fielded, cleaned
of cosmic rays, corrected for geometrical distortion and flux
calibrated. The total integration time was splitted in two or more
individual exposures.  In some objects, the telescope was offset by
several pixels along the slit direction between repeated exposures to
aid in the removal of hot pixels.  If there are more than two
exposures of the same object, the median frame was obtained.  For
those galaxies with two individual exposures, we combined the 2D
spectra using a statistical differencing technique similar to that
used by Pogge \& Martini (2002) for WFPC2 images. The technique
consists in obtaining for each individual exposure a mask frame
containing only the hot pixels and cosmic-ray hits, which is then
subtracted from the original spectrum. The mask is obtained as
follows. First, for each pair of exposures, a difference spectrum is
formed by subtracting one spectrum in the pair from the other. The
resulting frames consist mainly of positive and negative cosmic-rays
and hot pixels. A pair of mask frames is formed by separating the
remaining positive and negative pixels, setting all the pixels within
$\pm$ a few $\sigma$ of the mean residual background level on the
difference spectrum to zero.  This technique works well when the pair
of original spectra have similar spatial distribution along the slit,
and background.

All the spectra have been treated as extended continuum sources when
converting from surface brightness units in the two-dimensional frames
to flux units ({\ergsacm}). Finally, the spectra have been corrected
for redshift assuming the heliocentric radial velocities listed in
Table 1.
  
\begin{figure}
\caption{Distance and morphological type distributions for the strong-[OI]  
(lower pannel) and weak-[OI] (upper pannel) LLAGN
in the HFS97 sample. Filled areas in the histograms indicate the sample
analyzed here.}
\label{}
\end{figure}

\begin{deluxetable}{lccccccccc}
\tabletypesize{\scriptsize}
\tablewidth{0pc}
\tablecaption{LLAGN with archival STIS/CCD (G430L) spectra}
\tablehead{
\colhead{Name} &  \colhead{Type} &   \colhead{Morph} &  \colhead{v}    
 &  \colhead{distance} & \colhead{1 arcsec} & \colhead{[OI]/H$\alpha$} & \colhead{HST-ID} & \colhead{Exposure} & \colhead{P.A.} \ \nl
 \colhead{}  &  \colhead{}  &  \colhead{}  & \colhead{km/s} & \colhead{Mpc}  & \colhead{pc$^2$}  
& \colhead{} & \colhead{}  & \colhead{s}  & \colhead{$^{\circ}$} \nl
}
\startdata
NGC 2685$^*$& S2/T2&S0$_3$(7) pec&869 & 16.2 & 78  & 0.13b & 8607 & 2585 & 54.4 \nl 
NGC 2787 & L1.9  & SB(r)0+     & 691  & 13.0 & 63  & 0.55b & 7361 & 1864 & 213.2 \nl
NGC 3368 & L2    & SAB(rs)ab  & 897   & 8.1  & 39  & 0.18  & 7361 & 1574 & 249.5 \nl 
NGC 3489 & T2    & SAB(rs)0+  & 701   & 6.4  & 31  & 0.11b & 7361 & 1644 & 239.1 \nl
NGC 3507 & L2    & SB(s)b    &  978   & 19.8 & 96  & 0.18  & 7357 & 900  & 89.9 \nl
NGC 3627$^*$& T2/S2 & S(s)bII.2 & 703 & 6.6  & 32  & 0.13  & 8607 & 2349 & 80.1 \nl
NGC 3675 & T2    & SA(s)b     & 766   & 1.8  & 9   & 0.12b & 8607 & 2472 & 205.9 \nl
NGC 3953 & T2    & SB(r)bc    & 1053  & 17   & 82  & 0.12b & 8607 & 2561 & 79.1 \nl
NGC 3992 & T2:    & SB(rs)bc   & 1048 & 17   & 82  & 0.13c & 7361 & 1796 & 155.3 \nl
NGC 3998 & L1.9  &  SA0        & 1049 & 21.6 & 105 & 0.53  & 8839 & 2827 & 304.3 \nl
NGC 4143 & L1.9  & SAB(s)0     & 783  & 17   & 82  & 0.71  & 7361 & 1707 & 139.0 \nl
NGC 4150$^*$ & T2& S0$_3$(4)/a & 244  & 9.7  & 47  & 0.13  & 8607 & 2395 & 240.1\nl
NGC 4203 & L1.9   & SAB0-      & 1085 & 9.7  & 47  & 1.22b & 7361 & 1630 & 105.6 \nl
NGC 4261 & L2    & E2+         & 2210 & 35.1 & 170 & 0.49  & 8236 & 1891 & 157.9 \nl
NGC 4314 & L2    & SB(rs)a     & 962  & 9.7  & 47  & 0.18b & 7361 & 1668 & 105.3 \nl
NGC 4321 & T2    & SAB(s)bc    & 1234 & 16.8 & 81  & 0.11  & 7361 & 1642 & 92.9 \nl
NGC 4414 & T2:     & SA(rs)c    & 719 & 9.7  & 47  & 0.14u & 8607 & 2395 & 125.1 \nl
NGC 4429 & T2     & SA(r)0+    & 1137 & 16.8 & 81  & 0.097u& 8607 & 2349 & 81.1 \nl
NGC 4435 & T2    & SB(s)0      & 781  & 16.8 & 81  & 0.13b & 7361 & 1644 & 89.6 \nl 
NGC 4450 & L1.9  & SA(s)ab     & 1956 & 16.8 & 81  & 0.67  & 7361 & 1669 & 233.0 \nl
NGC 4459 & T2    & SA(r) 0+    & 1202 & 16.8 & 81  & 0.13u & 7361 & 1644 & 92.9 \nl
NGC 4548 & L2    & SB(rs)b     & 485  & 16.8 & 81  & 0.23  & 7361 & 1644 &  73.2 \nl
NGC 4569$^*$ & T2& S0$_3$(3)   & -311 & 16.8 & 81  & 0.062 & 8607 & 2349 &  100.0\nl
NGC 4596  & L2:: & SB(r)0+     & 1874 & 16.8 & 81  & 0.27u & 7361 & 1671 & 250.3\nl
NGC 4826$^*$& T2 &(R)SA(rs)ab  & 411  & 4.1  & 20  & 0.073 & 8607 & 2356 & 88.1 \nl
NGC 5055$^*$& T2 & S(s)bcII-III& 516  & 7.2  & 35  & 0.17u & 7361 & 1707 & 164.5 \nl
NGC 6503$^*$& T2/S2 & S(1)cII.8& 26   & 6.1  & 30  & 0.08  & 8607 & 2687 & 315.1 \nl
NGC 7331$^*$  & T2  & S(rs)bI-II& 835 & 14.3 & 69  & 0.097u& 8607 & 2395 & 358.9\nl
\hline
NGC 278  & H  & SAB(rs)b    & 639    & 11.8  & 57  & 0.022b& 7361 & 1786 & 47.6 \nl
NGC 3351 & H  & SB(r)b      & 778    & 8.1   & 39  & 0.019 & 7361 & 1671 & 245.6 \nl
NGC 4245 & H  & SB(r)0/a    & 890    & 9.7   & 47  & 0.038u& 7361 & 1668 & 85.7 \nl
NGC 4800 & H  & SA(rs)b     & 808    & 15.2  & 74  & 0.041b& 7361 & 1739 & 177.5\nl
NGC 1023$^*$& Abs& SB(rs)0     & 632    & 10.5  & 51& ...  & 7566 & 2475 & 93.1 \nl
\enddata
\tablenotetext{1}{Galaxies labeled with * have been observed also at NOT.}
\tablecomments{Col.\ (1): Galaxy name; Cols.\ (2): Spectral class (according to HFS97 criteria);
Cols.\ (3): Hubble type; Cols.\ (4): Radial velocity; Cols.\ (5): Distance; Cols.\ (6): Angular scale;
Cols.\ (7): [OI]/H$\alpha$ flux ratio; Cols.\ (8): HST proposal ID number; Cols.\ (9): Exposure time;
Cols.\ (10): Slit P.A.}
\end{deluxetable}

\section{Results}

\subsection{Central morphology and Nuclear spectral extractions}

To extract the nuclear spectra of these galaxies, we have examined the
spatial distribution of the continuum at 4700 \AA\ along the slit. In
all the objects, the central surface brightness is extended, with 
a symmetric distribution about the maximum, in many of them, this indicates
that the bulge component dominates the central continuum light. LLAGN
that have this type of distribution are: NGC 2787, NGC 3489, NGC 3992,
NGC 4429, NGC 4450, NGC 4596, and NGC 7331. A few of the galaxies show
a very sharp distribution that contains most of the central
flux. These are NGC 3998, NGC 4321 and NGC 4569. While the sharp
brightness distribution in NGC 4321 and NGC 4569 could be produced by
a nuclear stellar cluster, the origin in NGC 3998 is not clear.  Other
objects, like NGC 3627, NGC 4150, NGC 4261, NGC 4435, NGC 4548, and NGC
6503, show a central complex structure, that could be produced by the
superposition of dust lanes and several clusters randomly distributed
or in a circumnuclear ring. To better inspect the central morphology
of these LLAGN we have retrieved the WFPC2 optical images available in
the HST archive. These images confirm the morphology suggested by the
spectral central surface brightness profile. A detailed analysis of
the central morphology and its relation with the nuclear stellar
population is being carried out (Gonz\'alez Delgado, in preparation).
Figure 2 shows images and slit profiles of a few representative
examples.

We have extracted two spectra for each galaxy (except in the case of
NGC 4150, for which we have extracted three spectra, see Figure 3)
corresponding to the central 0.3 arcsec (called {\it b} spectra) and 1
arcsec (called {\it a} spectra). These two extractions allow us to
check if there is a change of the stellar population on a spatial
scale of a few tens of parcsecs. A direct comparison of the spectra of
these two extractions indicates that the dominat stellar population on
scales of 1 arcsec is also the dominant stellar population on the 0.3
arcsec scale.

Eight TOs (NGC 2685, NGC 3627, NGC 4150, NGC 4569, NGC 4826, NGC 5055,
NGC 6503, NGC 7331) and NGC 1023 have also been observed by us from
the ground.  We have compared the {\it a} spectra with the nuclear
spectra obtained from the ground. The strongest variations are found
in NGC 4569 and NGC 4150, with a significant change in the shape of
the continuum and/or the strength of the absorption lines. In the case
of NGC 4569, the ground-based nuclear spectrum is somewhat redder and
presents stronger metallic lines (such as the G band and CaII K) than
the STIS {\it a} spectrum, indicating that a larger fraction of the
bulge population contributes to the ground-based extraction
(1$\times$1.1 arcsec) compared to the STIS extraction (1$\times$0.2
arcsec). In the case of NGC 4150 the differences in the shape of the
nuclear spectra can be associated with the dust lane structure that
crosses the center and the different P.A. of the slit in the
ground-based observation with respect to the STIS observations. In the
remaining objects, the STIS and ground-based nuclear spectra are visually
similar (see, e.g. NGC 3627 in Figure 4 on this paper and Figure 8 on Paper I),
although some differences on the stellar populations are obtained through 
the empirical population analysis dicussed in section 4.

\begin{figure}
\caption{{\it Top:} WFPC2 images of NGC 3507, NGC 3627, and NGC 4826
through the filters F606W or F547M. North is up and East to the
left. {\it Bottom:} Surface brightness
profiles along the STIS slit of the optical continuum at 4700 \AA. The
extension of the {\it a} and {\it b} spectral extractions are marked
by horizontal lines and labeled.}
\label{}
\end{figure}

\begin{figure}
\caption{WFPC2 image through the F555W filter of NGC 4150, surface
surface brightness profile along the STIS slit of the optical
continuum at 4700 \AA, and STIS spectra compared with the ground-based
nuclear spectrum. The spectra are normalized to the flux at 4800 \AA\
and shifted vertically for clarity.}
\label{}
\end{figure}

\subsection{Empirical stellar population classification}

In Paper I LLAGN are classified empirically by comparing their spectra
with those of non-active galaxies dominated by stellar population of
different ages.  The spectra were separated into four classes labeled
by $\eta = $ {\it Y}, {\it I}, {\it I/O} and {\it O}:

\begin{itemize}

\item{{\it Y}: Galaxies with young ($\leq 10^{7}$ yr) stellar
population, characterized by a blue continuum and very diluted
metallic absorption lines.}
\item{{\it I}: Galaxies with a dominant intermediate age
population ($10^8$--$10^9$ yr), characterized by prominent HOBLs in
absorption.}
\item{{\it I/O}: Galaxies with a mixture of intermediate age and older
stars.  These galaxies do not have visible HOBLs in absorption, but
the metallic lines are weaker than in an old stellar population. }
\item{{\it O}: Galaxies dominated by an old ($10^{10}$ yr) stellar
population, with strong metallic lines, such as CaK and H, CN, G band
and MgII. }

\end{itemize}

Here, we proceed in the same way comparing the STIS spectra with the
template galaxies observed from the ground. Figures 4 to 7 group the
weak and strong-[OI] LLAGN spectra in the I, I/O and O categories.  As
in Paper I, this classification is confirmed by modelling of the
starlight by a combination of a base of five template galaxies
representative of the O (NGC 1023 and NGC 2950), I/O (NGC 221), I (NGC
205) and Y (NGC 3367) classes. The results of this combination indicate
that all the nuclei that belong to the O have relative contribution at 
4020 \AA\ of NGC 1023$+$NGC2950 larger than 75\%, and those that belong to
the I class have relative contribution at 4020 \AA\ of NGC 205 larger than
30\%. These results are in perfect agreement
with the those obtained from the modelling of the stellar light using
the population synthesis technique presented in section 4.

As in the ground-based spectra, none of the STIS LLAGN resembles pure
young stellar systems. No features due to WR stars were
detected directly in the STIS spectra of either LINERs or TOs, in
neither extraction {\it b} (0.3$\times$0.2 arcsec) nor {\it a}
(1$\times$0.2 arcsec). If WR stars are present in the nuclei of some
LLAGN, they must contribute very little to the optical light, given
that even observing through a very narrow slit, thus minimizing the
bulge light, the broad bump at 4660 \AA\ is not detected. This result
is also maintained when the spectra are subtracted of the intermediate
and old stellar components (obtained through the template
decomposition discussed above). Only in NGC 3507 we find a marginal
evidence of a broad feature at 4620 \AA\ in the residual spectrum,
although no broad feature at 4680 \AA\ is detected. 
  
Most of the LLAGN spectra are dominated by starlight, with weak or
absent nebular emission. Three LINERs (NGC 3998, NGC 4203,
and NGC 4450), which are classified as L1.9 for presenting 
broad Balmer lines,
 are the exception (Figure 7b). These objects are among
the strongest [OI]/H$\alpha$ emitters of the HFS97 sample, with values 
of 0.53, 1.22, and 0.67, respectively. Another important feature of 
these objects is that the
stellar lines are very diluted with respect to the strength of these
lines in a typical old and/or intermediate age population. This
resembles what is observed in many Seyfert 2 galaxies (Schmitt et al
1999; Gonz\'alez Delgado et al 2001; Cid Fernandes et al 2001a) in
which the optical continuum is modelled by an old age
population plus a feature less continuum (fitted by a power-law). 
In section 5 we discuss the
origin of the continuum in these LINERs. None of the weak-[OI] LLAGN 
observed have such properties.

We find that an intermediate age population is more frequent in TOs
(7/17) than in LINERs (2/11). Note, however, that both NGC 3368 and
NGC 3507, the two $\eta = I$ LINERs, have [OI]/H$\alpha = 0.18$,
therefore tresspassing the LINER/TO boundary of HFS97c by an
insignificant amount. Expressed in terms of our weak and strong-[OI]
classes (divided at [OI]/H$\alpha = 0.25$), this difference becomes
even more pronounced: 9/21 weak-[OI] but {\it none} of the strong-[OI]
nuclei in the STIS sample present HOBLs. Taking these numbers together
with the data from Paper I, we find that only 2 (NGC 841,
[OI]/H$\alpha$=0.58, and NGC 5005, [OI]/H$\alpha$=0.65) out of the 24
(8\%) strong-[OI] LLAGN belong to the $\eta =$ {\it I} class, while
this fraction is 20/47 (42\%) for weak-[OI]. As pointed out in Paper
I, these numbers suggest a possible link between the central stellar
population and emission line properties.

Figure 8 shows the STIS spectra of four HII nuclei (NGC 278, NGC 3351,
NGC 4245, and NGC 4800) and one non-active galaxy (NGC 1023) to be
compared with the LLAGN. Paradoxically, not all the HII nuclei show a
continuum dominated by young and/or intermediate age population!  In
fact, the dominant central $0.3 \times 0.2$ arcsec stellar population
in NGC 4245 and NGC 4800 is old ($\ga 10^{10}$ yr) and somewhat
similar to the non-active galaxy NGC 1023. However, the H$\alpha$
distribution in both objects is extended and more prominent in the
off-nuclear spectra than in the central 0.3$\times$0.2 arcsec, where
H$\alpha$ is in absorption. It would be interesting to revise the
emission line classification of these objects, and to check if their
HII denomination is associated to the large aperture of the spectra
used in the HFS97 classification; then it is associated to star formation 
located in the circumnuclear region (100 pc to 1 kpc) than to the nucleus.
 This comparison also indicates the
relevance of the central morphology of the continuum and emission
lines in the classification of LLAGN.

\begin{figure}
\begin{center}
\end{center}
\caption{Nuclear spectra of weak-[OI] in the {\it I} class, i.e., with
an intermediate age population.}
\label{}
\end{figure}

\begin{figure}
\begin{center}
\end{center}
\caption{Nuclear spectra of weak-[OI] in the {\it I/O} class, i.e.,
with a mixture of intermediate age and older stars. Hot pixels in the spectra are labbeled by *. }
\label{}

\end{figure}

\begin{figure}
\begin{center}
\end{center}
\caption{Nuclear spectra of weak-[OI] in the {\it O} class, i.e., with
a predominantly old age population.}
\label{}
\end{figure}

\begin{figure}
\begin{center}
\end{center}
\caption{Nuclear spectra of strong-[OI]: (a) in the {\it O} class,
i.e., with a predominantly old age population; (b) with diluted
stellar lines }
\label{}
\end{figure}

\begin{figure}
\begin{center}
\end{center}
\caption{Nuclear spectra of a non-active galaxy, NGC 1023, and four
HII nuclei. }
\label{}
\end{figure}

\subsection{Spectral properties}

In this section we present measurements of spectral indices indicative
of stellar populations for the STIS spectra. We list the measurements
only for extractions {\it a}, since in most of the objects the
differences in the spectral indices with respect to extractions {\it
b} are small. We have measured 7 equivalent widths and two continuum
colors in the system of Bica \& Alloin (1986a,b): $W_C$, $W_{wlb}$,
$W_K$, $W_H$, $W_{CN}$, $W_G$, $W_{Mg}$, $F_{3660}/F_{4020}$ and
$F_{4510}/F_{4510}$. These measurements were performed following the
automated procedure devised in Paper I except for 3 cases (NGC 4459,
NGC 4569 and NGC 6503), where the pseudo continuum was corrected by
hand. We have also measured the 4000 \AA\ break index ($D_n(4000)$),
and the [OII] and H$\delta$ (H$\delta_A$) equivalent widths following
recipes in Balogh et al (1999) and Worthey \& Ottaviani (1997).

Table 2 lists the results of these
measurements. In Figure 9 we examine the relation between $D_n(4000)$,
H$\delta_A$ and $W_K$ for the combined STIS + ground samples.
Different symbols correspond to stellar population classes $\eta = $
I, I/O and O (circles, triangles and squares respectively). Filled and
open symbols are used for STIS and ground-based observations
respectively. As discussed in Paper I, there is an excellent agreement
between the empirical stellar population classes and the spectral
indices. Nuclei with young and/or intermediate age populations ($\eta
=$ {\it I}) have low $W_K$ and $D_n(4000)$, while for nuclei dominated
by an old stellar population these indices assume large values, with
$\eta =$ {\it I/O} objects in between.

\begin{figure}
\begin{center}
\end{center}
\caption{Relations between spectral indices: D$_n$(4000), H$\delta_A$
and $W_K$. The stellar population classes {\it I, I/O, O} are
represented by different symbols (circles, triangles and squares
repectively). Except for the two stars, which represent the HII nuclei
NGC 3367 and NGC 6217 from Paper I, only LLAGN are included. Crosses
are used to represent NGC 3998, NGC 4143, NGC 4203 and NGC 4450, the
four STIS sources with strong emission lines (Fig 7b). Filled and open
symbols are used for STIS and ground-based observations
respectively.}
\label{}
\end{figure}

\begin{deluxetable}{lrrrrrrrrrrrr}
\tabletypesize{\scriptsize}
\tablewidth{0pc}
\tablecaption{Spectral properties in the {\it a} LLAGN STIS/CCD (G430L) spectra}
\tablehead{
\colhead{Galaxy}        &
\colhead{$W_C$}         &
\colhead{$W_{wlb}$}     &
\colhead{$W_K$}         &
\colhead{$W_H$}         &
\colhead{$W_{CN}$}      &
\colhead{$W_G$}         &
\colhead{$W_{Mg}$}      &
\colhead{$\frac{F_{3660}}{F_{4020}}$} &
\colhead{$\frac{F_{4510}}{F_{4020}}$} &
\colhead{$D_n(4000)$}     & 
\colhead{H$\delta_A$}     &
\colhead{$W$(\ion{O}{2})} }
\startdata
NGC 2685 &   5.2 &  20.5 &  18.9 &  13.7 &  15.8 &   9.8 &  11.0 &  0.57 & 1.62 &    2.17 &   -1.6 &    2.1 \nl
NGC 2787 &   5.9 &  20.6 &  16.8 &  11.3 &  19.2 &  10.7 &  11.4 &  0.54 & 1.53 &    2.08 &   -1.0 &   26.1 \nl
NGC 3368 &   1.7 &   9.7 &  10.1 &  11.8 &   8.3 &   6.9 &   6.8 &  0.53 & 1.23 &    1.48 &    3.8 &    1.2 \nl
NGC 3489 &   2.2 &  12.2 &  13.0 &  12.3 &   8.8 &   7.9 &   7.4 &  0.56 & 1.16 &    1.58 &    3.5 &   -2.0 \nl
NGC 3507 &   1.0 &   4.1 &   5.4 &   6.6 &   4.6 &   4.6 &   4.9 &  0.81 & 1.03 &    1.26 &    0.9 &    6.5 \nl
NGC 3627 &   1.5 &   9.7 &   9.5 &  11.1 &   5.1 &   6.3 &   6.1 &  0.47 & 1.36 &    1.54 &    4.9 &   11.2 \nl
NGC 3675 &   2.3 &  18.6 &  18.4 &  13.2 &  16.3 &  10.4 &  10.9 &  0.60 & 1.71 &    2.17 &   -2.0 &   -2.4 \nl
NGC 3953 &   2.4 &  12.4 &  13.1 &  11.9 &  11.4 &   8.5 &   8.5 &  0.50 & 1.62 &    1.84 &    0.6 &    4.2 \nl
NGC 3992 &   3.3 &  19.0 &  18.7 &  12.7 &  17.6 &  10.5 &  11.1 &  0.46 & 1.67 &    2.36 &   -1.4 &    6.2 \nl
NGC 3998 &   1.2 &  -1.6 &   3.9 &  -6.1 &  12.7 &   0.3 &   4.8 &  0.93 & 1.25 &    1.25 &   -3.1 &   50.2 \nl
NGC 4143 &   3.4 &  14.2 &  14.8 &   8.2 &  17.4 &   8.7 &  11.5 &  0.63 & 1.49 &    1.72 &   -1.1 &   22.6 \nl
NGC 4150 &   2.0 &   9.5 &   7.9 &  10.4 &   5.5 &   7.4 &   5.8 &  0.54 & 1.38 &    1.50 &    3.5 &    5.5 \nl
NGC 4203 &   2.4 &   9.2 &  10.0 &   3.2 &  17.4 &   6.6 &   8.9 &  0.77 & 1.40 &    1.42 &   -2.6 &   19.2 \nl
NGC 4261 &   4.4 &  18.9 &  18.4 &  12.7 &  18.1 &   9.5 &  10.2 &  0.61 & 1.55 &    1.92 &   -4.4 &   12.0 \nl
NGC 4314 &   6.3 &  18.2 &  15.1 &  14.3 &  15.3 &  10.7 &  10.7 &  0.47 & 1.60 &    2.02 &   -2.5 &   -2.8 \nl
NGC 4321 &  -0.8 &   5.4 &   7.5 &   7.3 &   9.5 &   6.5 &   6.8 &  0.70 & 1.23 &    1.32 &    1.2 &    5.8 \nl
NGC 4414 &   3.6 &  16.5 &  16.6 &  12.7 &  14.7 &   9.8 &   9.4 &  0.55 & 1.42 &    1.91 &   -0.9 &    2.1 \nl
NGC 4429 &   4.9 &  19.8 &  15.8 &  13.4 &  18.5 &  10.4 &  12.2 &  0.52 & 1.60 &    1.95 &   -1.7 &   -1.8 \nl
NGC 4435 &   5.1 &  16.1 &  15.1 &  13.2 &  12.2 &   9.1 &  10.9 &  0.43 & 1.38 &    1.96 &   -1.2 &   -1.7 \nl
NGC 4450 &   3.7 &  11.0 &  11.7 &   6.4 &  14.5 &   7.9 &   9.0 &  0.74 & 1.42 &    1.52 &   -1.6 &   30.2 \nl
NGC 4459 &   4.5 &  18.3 &  14.8 &  11.4 &  15.5 &  10.1 &  11.9 &  0.62 & 1.78 &    2.21 &   -5.3 &   -2.3 \nl
NGC 4548 &   6.2 &  23.1 &  19.0 &  12.1 &  18.5 &  10.2 &  11.7 &  0.40 & 1.76 &    2.36 &   -5.4 &   10.5 \nl
NGC 4569 &   0.2 &   4.6 &   3.4 &   7.5 &   3.2 &   2.9 &   2.9 &  0.66 & 1.04 &    1.13 &    5.7 &    0.5 \nl
NGC 4596 &   4.0 &  17.0 &  15.9 &  11.6 &  15.9 &   9.7 &  11.0 &  0.55 & 1.56 &    2.05 &   -1.7 &   -1.0 \nl
NGC 4826 &   2.5 &  13.7 &  12.9 &  11.9 &  12.0 &   8.3 &   8.5 &  0.52 & 1.37 &    1.70 &    0.7 &    2.8 \nl
NGC 5055 &   3.0 &  12.9 &  11.9 &  10.3 &  13.9 &   7.1 &  10.9 &  0.63 & 1.48 &    1.60 &   -0.4 &   -1.7 \nl
NGC 6503 &   3.3 &   4.6 &   6.5 &   9.6 &   4.4 &   4.6 &   5.8 &  0.69 & 1.12 &    1.27 &    3.8 &   -1.5 \nl
NGC 7331 &   4.1 &  16.4 &  17.9 &  13.6 &  16.2 &  10.2 &   9.7 &  0.57 & 1.53 &    2.02 &   -1.4 &   -1.6 \nl
\hline
NGC 0278 &   0.2 &   7.8 &   4.3 &  12.2 &   3.9 &   4.3 &   3.5 &  0.39 & 0.90 &    1.31 &   10.1 &   -1.6 \nl
NGC 1023 &   5.7 &  21.5 &  19.3 &  14.0 &  21.6 &  11.6 &  12.6 &  0.57 & 1.61 &    2.32 &   -3.9 &   -4.3 \nl
NGC 3351 &   2.4 &  10.0 &   9.2 &   8.4 &   4.9 &   6.6 &   5.5 &  0.70 & 1.31 &    1.44 &   -0.5 &   -4.2 \nl
NGC 4245 &   3.8 &  16.4 &  14.4 &  11.0 &  16.7 &  11.0 &  11.1 &  0.58 & 1.67 &    1.93 &   -3.0 &   -2.8 \nl
NGC 4800 &   5.0 &  16.9 &  16.9 &  11.0 &  13.0 &   8.2 &   8.5 &  0.62 & 1.55 &    1.93 &   -2.1 &   -2.9 \nl
\enddata
\label{tab:SpecIndices}
\tablecomments{Col.\ (1): Galaxy name; Cols.\ (2--10): Equivalent
widths of seven absorption features and two colors, all in Bica \& Alloin 
system. Col.\ (11): 4000 \AA\ break index (Balogh et al 1999).  Col.\
(12): H$\delta_A$ equivalent width of Worthey \& Ottaviani (1997).
Col.\ (13): [OII] equivalent width of Balogh et al (1999).}
\end{deluxetable}

\section{Empirical Population Synthesis Analysis}

\subsection{The method}

In order to quantify the stellar population mixture of LLAGN we use
the empirical population synthesis (EPS) algorithm described in Cid
Fernandes et al (2001b). Briefly, the code synthesizes a set of
equivalent widths and colors by means of a combination of a base of 12
observed star clusters of differents ages and metallicities (Schmidt
et al 1991; Bica \& Alloin 1986a,b) plus a $F_\nu \propto \nu^{-1.5}$
power-law to represent an AGN featureless continuum (FC). The output
of the code consists of a {\it population vector} $\vec{x}$, whose
components represent flux fractions associated with each population in
the base, plus the V-band extintion, modelled as due to a uniform dust
screen.  These parameters correspond to a likelyhood-weighted mean of
$10^8$ combinations obtained from a Metropolis tour through the
($\vec{x},A_V$) space. We use as input to the EPS code five spectral
indices: $W_K$, $W_{CN}$, $W_G$ plus the F$_{3660}$/F$_{4020}$ and
F$_{4510}$/F$_{4020}$ colors.  This same set of observables was used
in the EPS analysis of Starburst (Cid Fernandes, Le\~ao \& Rodrigues
Lacerda 2003b) and Seyfert 2 galaxies (Cid Fernandes \etal 2001a),
which provide an important reference for comparison.

The virtues and shortcomings of EPS have been extensively discussed in
Cid Fernandes \etal (2001b, 2003b). Our experience with this method in
a variety studies has taught us that it is a superb tool to analyse
stellar population mixtures provided that one keeps the description
at a relatively coarse level.  Accordingly, in this paper we group
components of same age in a reduced population vector $\vec{x} =
(x_{FC},x_6, x_7,x_8,x_9,x_{10})$ where $x_6$--$x_{10}$ correspond to
five logarithmically spaced ages of $10^6$, $10^7$, $10^8$, $10^9$ and
$10^{10}$ yr, and $x_{FC}$ corresponds to the power-law component.
Even this description is too fine graded given the limitations of the
EPS imposed by the combination of observational errors, limited input
information and quasi-linear dependences within the base. We therefore
base our analysis of the EPS results on an even coarser (but more
robust) description obtained by further grouping similar $\vec{x}$
components. For instance, although the method does not distinguish
well between the $10^6$, $10^7$ and FC individual components, their
sum $x_{Y/FC} \equiv x_{FC} + x_6 + x_7$ is well constrained. Similar
rebinnings of the remaining components are employed in the discussion
below.

We have also calculated formal errors and covariances for
three condensed EPS fractions (x$_{Y/FC}$, x$_8$ and x$_{9+10}$).
The average errors are 2.2$\%$, 3.2$\%$, and 3.2$\%$, respectively.
These errors are low, and allow us to distinguish between the different
components with a precision of about 3$\%$. However, this does not 
apply for old systems with a contribution of x$_{9+10}\geq$ 90$\%$.
In these objects, the fractions x$_{Y/FC}$ and x$_8$ obtained are 
similar to the formal errors, and in consequence these values are not
significant.

\subsection{Statistical results}

Tables 3 and 4 list the EPS results for both the ground-based and STIS 
samples. The population vector is expressed as the percentage fraction
of the flux at a normalization wavelength of 4020 \AA.  In this
section we present histograms of the results of the synthesis grouped
in three age bins, used to represent old ($10^{10}$ yr, $x_{10}$),
intermediate populations ($10^8$--$10^9$ yr, $x_8 + x_9$) and young
plus FC components ($x_{Y/FC}$). In these histograms we divide the
LLAGN in weak-[OI] ([OI]/H$\alpha \leq$ 0.25) and strong-[OI]
([OI]/H$\alpha \geq$ 0.25) sources.

Figure 10 shows the results for the
ground-based observations only.  The average (median) values of the
distributions of the old, intermediate and young stellar
population of the weak-[OI] LLAGN are: 48\% (47\%), 46\% (45\%) and
6\% (5\%), repectively.  For the strong-[OI] LLAGN these values are:
62\% (64\%), 35 \% (33\%) and 3\% (3\%).  A remarkable result from
this analysis is that the $Y/FC$ component rarely exceeds 10\% of the
flux in the 1.1$\times$1 arcsec ground-based aperture. This only
happens in 5/51 (10\%) of our LLAGN (NGC 772, NGC 4569, NGC 5377, NGC
5678 and NGC 6503), all of which are weak-[OI] sources. Another
significant difference between strong and weak-[OI] is found in the
fraction of intermediate age population.  The fraction of weak-[OI]
LLAGN with $x_I=x_8 + x_9 > 35\%$ is  25/34 (74\%), but only 5/17 (29\%) for
strong-[OI]. Regarding the old stellar population, the fraction of
weak-[OI] nuclei with $x_{10} < 65\%$ is 27/34 (79\%) and  10/17
(59\%) for strong-[OI].  These numbers essentially translate the
analysis done in Paper I to population strengths, and clearly indicate
that weak-[OI] LLAGN are nuclei with an important intermediate age
population.

\begin{figure}
\caption{Histograms of the contribution of different age components,
normalized to the light at 4020 \AA. The histogram is for the
ground-based observations, thus corresponding to stellar populations
that contribute to a nuclear aperture of 1.1$\times$1 arcsec.  The
filled area represents the distributions of weak-[OI] LLAGN. Labels W
and S indicate the median corresponding to the distributions of weak
and strong-[OI] LLAGN respectively. }
\label{}
\end{figure}

Figure 11 shows the results for the STIS {\it a} spectra.  The EPS
analysis has also been done for the {\it b} spectra, but we have not
found a significant difference between the results for the two
apertures. The previous conclusion that weak-[OI] LLAGN have a larger
contribution from the intermediate age population than strong-[OI]
LLAGN is still valid. In 16 of the 21 (76\%) weak-[OI] but only 2/7
(28\%) of the strong-[OI] LLAGN the contribution of the intermediate
age population is larger than 35\%.  However, there is a noticeable
difference with respect to the ground-based observations, related with
the strength of young plus FC component. The mean value of this
component is $x_{Y/FC} = 12\%$ and 22\% for weak and strong-[OI] LLAGN
respectively. These values are larger than the mean contributions in
the ground-based spectra, 3\% and 6\%, respectively.  There are four
weak-[OI] (NGC 3507, NGC 4321, NGC 4569, and NGC 6503) and three
strong-[OI] LLAGN (NGC 3998, NGC 4203, and NGC 4450) with $x_{Y/FC} >
20\%$, whereas only NGC 4569 has such a strong component in the ground
based data. These three LINERs show very strong emission lines and
diluted stellar lines, probably produced by the contribution from an
AGN component (see section 5.1).  On the other hand, in NGC 3507, NGC
4321, NGC 4569 and NGC 6503, the UV emission and the spectral
characteristics of the optical spectra suggest that the large
contribution from the $Y/FC$ component is provided by a young stellar
cluster.  This cluster must be compact, as indicated by the morphology
in the WFPC2 images and surface brightness profile along the slit, and
hence its contribution to the optical light is much less important
into a ground-based aperture. It is interesting to note that in the
ground-based aperture, NGC 4569 and NGC 6503 outstand for their large
intermediate age population.

\begin{figure}
\caption{As Figure 10 but for STIS {\it a} spectra (corresponding to
the nuclear 1$\times$0.2 arcsec). }
\label{}
\end{figure}

\begin{deluxetable}{lccccccc}
\tabletypesize{\tiny}
\tablewidth{0pc}
\tablecaption{EPS results for the ground-based observations}
\tablehead{
\colhead{Galaxy}    &  
\colhead{$x_{10}$}  &
\colhead{$x_9$}     &
\colhead{$x_8$}     &
\colhead{$x_7$}     & 
\colhead{$x_6$}     & 
\colhead{$x_{PL}$}  & 
\colhead{$A_V$}     }
\startdata
NGC 0266      &   63.3 &  33.0 &   1.8 &   1.0 &   0.4 &   0.4 &     0.28 \nl
NGC 0315      &   68.2 &  23.3 &   3.6 &   2.6 &   1.2 &   1.1 &     0.34 \nl
NGC 0404      &   27.3 &  35.7 &  28.7 &   4.4 &   1.9 &   1.9 &     1.56 \nl
NGC 0410      &   65.6 &  27.7 &   3.1 &   1.9 &   0.8 &   0.8 &     0.30 \nl
NGC 0428      &   25.6 &  64.4 &   6.3 &   2.1 &   0.8 &   0.9 &     0.79 \nl
NGC 0521      &   61.4 &  33.4 &   2.7 &   1.3 &   0.6 &   0.6 &     0.56 \nl
NGC 0660      &   42.0 &  48.4 &   5.1 &   2.5 &   1.0 &   1.1 &     2.37 \nl
NGC 0718      &   36.3 &  46.1 &  11.6 &   3.3 &   1.4 &   1.4 &     0.74 \nl
NGC 0772      &   46.9 &  24.4 &  11.5 &   8.7 &   4.9 &   3.5 &     0.28 \nl
NGC 0841      &   39.5 &  49.9 &   6.1 &   2.5 &   1.0 &   1.1 &     0.57 \nl
NGC 1052      &   70.1 &  20.8 &   3.9 &   2.8 &   1.2 &   1.1 &     0.37 \nl
NGC 1161      &   75.5 &  20.9 &   1.8 &   1.0 &   0.4 &   0.4 &     0.24 \nl
NGC 1169      &   73.2 &  22.9 &   1.7 &   1.2 &   0.5 &   0.5 &     0.15 \nl
NGC 2681      &   25.7 &  52.7 &  16.1 &   3.0 &   1.2 &   1.3 &     0.91 \nl
NGC 2685      &   65.2 &  31.3 &   1.8 &   1.0 &   0.4 &   0.4 &     0.26 \nl
NGC 2911      &   64.4 &  29.7 &   3.1 &   1.5 &   0.7 &   0.6 &     0.80 \nl
NGC 3166      &   42.9 &  48.3 &   4.9 &   2.1 &   0.9 &   0.9 &     0.62 \nl
NGC 3169      &   50.2 &  44.2 &   3.0 &   1.5 &   0.6 &   0.6 &     1.15 \nl
NGC 3226      &   78.3 &  16.4 &   2.4 &   1.5 &   0.7 &   0.6 &     0.53 \nl
NGC 3245      &   59.4 &  27.4 &   5.2 &   4.2 &   2.0 &   1.8 &     0.33 \nl
NGC 3627      &   29.6 &  47.1 &  16.7 &   3.6 &   1.4 &   1.5 &     1.14 \nl
NGC 3705      &   36.7 &  52.6 &   6.5 &   2.3 &   0.9 &   1.0 &     1.08 \nl
NGC 4150      &   29.2 &  52.3 &  12.7 &   3.2 &   1.2 &   1.3 &     0.62 \nl
NGC 4192      &   52.6 &  38.1 &   4.5 &   2.6 &   1.1 &   1.1 &     1.77 \nl
NGC 4438      &   67.2 &  27.2 &   2.8 &   1.5 &   0.7 &   0.6 &     1.15 \nl
NGC 4569      &   19.6 &  13.2 &  45.7 &  11.0 &   5.0 &   5.4 &     0.35 \nl
NGC 4736      &   36.8 &  43.9 &  12.6 &   3.7 &   1.6 &   1.5 &     0.68 \nl
NGC 4826      &   44.6 &  40.1 &   9.6 &   3.1 &   1.3 &   1.3 &     0.67 \nl
NGC 5005      &   34.4 &  53.4 &   7.4 &   2.6 &   1.0 &   1.1 &     1.38 \nl
NGC 5055      &   46.1 &  34.1 &  12.4 &   4.0 &   1.7 &   1.7 &     0.81 \nl
NGC 5377      &   31.1 &  24.3 &  32.5 &   6.2 &   3.1 &   2.8 &     0.64 \nl
NGC 5678      &   34.8 &  31.8 &  22.1 &   5.9 &   2.7 &   2.8 &     1.85 \nl
NGC 5879      &   33.7 &  60.5 &   3.3 &   1.4 &   0.5 &   0.6 &     0.89 \nl
NGC 5921      &   29.8 &  45.2 &  17.6 &   4.1 &   1.6 &   1.8 &     0.68 \nl
NGC 5970      &   60.7 &  35.8 &   1.6 &   1.0 &   0.4 &   0.5 &     0.18 \nl
NGC 5982      &   58.7 &  36.7 &   2.2 &   1.3 &   0.5 &   0.5 &     0.21 \nl
NGC 5985      &   65.0 &  31.4 &   1.7 &   1.0 &   0.4 &   0.5 &     0.22 \nl
NGC 6340      &   85.4 &  12.0 &   1.3 &   0.7 &   0.3 &   0.3 &     0.91 \nl
NGC 6384      &   69.1 &  27.4 &   1.4 &   1.1 &   0.5 &   0.5 &     0.13 \nl
NGC 6482      &   69.7 &  26.4 &   1.7 &   1.2 &   0.5 &   0.5 &     0.13 \nl
NGC 6500      &   60.8 &  26.4 &   3.6 &   4.9 &   2.4 &   1.8 &     0.12 \nl
NGC 6501      &   63.1 &  28.3 &   4.4 &   2.2 &   1.0 &   0.9 &     0.42 \nl
NGC 6503      &   39.3 &  31.4 &  12.6 &   8.7 &   3.9 &   4.1 &     0.43 \nl
NGC 6702      &   62.8 &  32.4 &   2.1 &   1.4 &   0.6 &   0.6 &     0.21 \nl
NGC 6703      &   71.3 &  24.3 &   2.1 &   1.2 &   0.5 &   0.5 &     0.31 \nl
NGC 6951      &   51.2 &  40.8 &   4.5 &   1.9 &   0.8 &   0.8 &     0.37 \nl
NGC 7177      &   60.5 &  32.4 &   3.4 &   2.0 &   0.9 &   0.9 &     0.57 \nl
NGC 7217      &   75.7 &  20.8 &   1.7 &   0.9 &   0.4 &   0.4 &     0.41 \nl
NGC 7331      &   61.8 &  33.3 &   2.5 &   1.3 &   0.6 &   0.6 &     0.43 \nl
NGC 7626      &   69.4 &  25.1 &   2.6 &   1.6 &   0.7 &   0.6 &     0.27 \nl
NGC 7742      &   52.3 &  41.6 &   3.2 &   1.5 &   0.6 &   0.7 &     0.64 \nl
\hline
NGC 3367      &    9.1 &   6.4 &   5.3 &  31.9 &  32.5 &  14.8 &     0.11 \nl
NGC 6217      &   11.2 &   6.7 &  34.2 &  16.5 &  17.4 &  13.9 &     0.26 \nl
NGC 0205      &   25.4 &  24.5 &  38.5 &   5.9 &   2.8 &   2.8 &     0.60 \nl
NGC 0221      &   56.4 &  39.1 &   2.1 &   1.3 &   0.6 &   0.6 &     0.20 \nl
NGC 0224      &   67.1 &  25.5 &   3.2 &   2.3 &   1.0 &   0.9 &     0.22 \nl
NGC 0628      &   55.8 &  37.4 &   2.9 &   2.0 &   0.8 &   0.9 &     0.35 \nl
NGC 1023      &   79.6 &  17.3 &   1.5 &   0.9 &   0.4 &   0.4 &     0.16 \nl
NGC 2950      &   60.0 &  33.4 &   3.1 &   1.9 &   0.8 &   0.8 &     0.23 \nl
NGC 6654      &   75.2 &  20.6 &   1.9 &   1.3 &   0.6 &   0.5 &     0.14 \nl
\enddata
\end{deluxetable}

\begin{deluxetable}{lccccccc}
\tabletypesize{\scriptsize}
\tablewidth{0pc}
\tablecaption{EPS results for the STIS {\it a} data}
\tablehead{
\colhead{Galaxy}    &  
\colhead{$x_{10}$}  &
\colhead{$x_9$}     &
\colhead{$x_8$}     &
\colhead{$x_7$}     & 
\colhead{$x_6$}     & 
\colhead{$x_{PL}$}  & 
\colhead{$A_V$}     }
\startdata
NGC 2685        &   66.4 &  29.4 &   2.0 &   1.2 &   0.5 &   0.5 &     0.51 \nl
NGC 2787        &   60.5 &  31.4 &   4.4 &   2.1 &   0.9 &   0.8 &     0.38 \nl
NGC 3368        &   33.1 &  31.7 &  24.8 &   5.5 &   2.5 &   2.4 &     0.73 \nl
NGC 3489        &   32.1 &  50.5 &  10.9 &   3.5 &   1.4 &   1.5 &     0.31 \nl
NGC 3507        &   26.8 &  14.4 &  14.7 &  17.4 &  15.2 &  11.5 &     0.25 \nl
NGC 3627        &   31.7 &  36.5 &  22.3 &   5.1 &   2.2 &   2.2 &     1.66 \nl
NGC 3675        &   74.1 &  20.3 &   2.5 &   1.7 &   0.7 &   0.7 &     0.59 \nl
NGC 3953        &   42.5 &  36.5 &  12.8 &   4.4 &   1.9 &   1.9 &     1.72 \nl
NGC 3992        &   59.1 &  37.4 &   1.9 &   0.9 &   0.4 &   0.4 &     0.91 \nl
NGC 3998        &   15.5 &   5.8 &   5.7 &  33.7 &  20.7 &  18.6 &     1.47 \nl
NGC 4143        &   57.0 &  24.7 &   8.5 &   5.3 &   2.4 &   2.1 &     0.43 \nl
NGC 4150        &   39.3 &  20.6 &  22.0 &   7.9 &   5.7 &   4.5 &     1.48 \nl
NGC 4203        &   48.9 &   8.6 &   9.0 &  15.3 &  11.9 &   6.3 &     0.51 \nl
NGC 4261        &   65.4 &  29.2 &   2.5 &   1.6 &   0.6 &   0.7 &     0.25 \nl
NGC 4314        &   51.7 &  34.8 &   8.6 &   2.6 &   1.2 &   1.1 &     1.10 \nl
NGC 4321        &   35.5 &  12.9 &  19.0 &  13.3 &  11.5 &   7.7 &     0.56 \nl
NGC 4414        &   48.0 &  44.0 &   4.2 &   2.0 &   0.8 &   0.8 &     0.44 \nl
NGC 4429        &   59.4 &  28.0 &   7.6 &   2.7 &   1.2 &   1.1 &     0.72 \nl
NGC 4435        &   27.8 &  61.6 &   7.1 &   1.9 &   0.8 &   0.8 &     1.17 \nl
NGC 4450        &   53.4 &  13.6 &   8.4 &  12.1 &   7.6 &   5.0 &     0.41 \nl
NGC 4459        &   65.4 &  17.2 &   6.3 &   5.6 &   3.1 &   2.3 &     1.25 \nl
NGC 4548        &   57.9 &  39.1 &   1.6 &   0.7 &   0.3 &   0.3 &     1.26 \nl
NGC 4569        &   14.6 &   9.1 &  40.9 &  13.2 &  11.1 &  11.0 &     0.79 \nl
NGC 4596        &   54.0 &  34.3 &   6.4 &   2.9 &   1.2 &   1.2 &     0.79 \nl
NGC 4826        &   39.6 &  37.0 &  15.9 &   4.1 &   1.7 &   1.7 &     0.82 \nl
NGC 5055        &   44.0 &  24.0 &  14.1 &   9.5 &   4.5 &   3.9 &     1.08 \nl
NGC 6503        &   28.7 &  19.3 &  21.4 &  14.3 &   8.1 &   8.2 &     0.72 \nl
NGC 7331        &   60.4 &  34.1 &   2.7 &   1.5 &   0.6 &   0.6 &     0.39 \nl
\hline
NGC 0278        &   11.8 &  10.1 &  71.1 &   3.4 &   1.8 &   1.8 &     0.40 \nl
NGC 1023        &   78.7 &  18.3 &   1.3 &   0.9 &   0.4 &   0.3 &     0.12 \nl
NGC 3351        &   45.3 &  22.6 &   9.3 &  10.7 &   6.5 &   5.7 &     0.95 \nl
NGC 4245        &   64.8 &  16.3 &   9.0 &   5.0 &   2.9 &   2.0 &     0.89 \nl
NGC 4800        &   55.4 &  36.4 &   3.6 &   2.5 &   1.0 &   1.1 &     0.81 \nl
\enddata
\end{deluxetable}

\section{Discussion}

\subsection{Comparison with non-active galaxies}

The EPS analysis suggests that many of the LLAGN have a $10^8$--$10^9$
yr intermediate age stellar population which is far more conspicous in
weak-[OI] than in strong-[OI] nuclei.  The
question arises whether these statistics are the typical stellar
population of early type galaxies, or intermediate age stars
are more significant in LLAGN than in non-active galaxies of the same
morphological type.

 Raimann et al (2003) have analyzed the stellar population of a small 
sample of S0 to Sbc non-active galaxies using the same EPS code and
star cluster base. They find that the old and the intermediate age
population contributes with $x_O$= 65$\%$ and $x_I$= 35$\%$ to the continuum at 4020 \AA.  
The mean contributions in our ground-based aperture are $x_O$=62$\%$ and
$x_I$=35$\%$ for strong-[OI] and $x_O$=48$\%$ and
$x_I=46\%$ for weak-[OI] LLAGN. On the other hand, 
the intermediate age population contributes with
$x_I \geq 35\%$ in 74$\%$ of the weak-LLAGN, while 
this happens in only 29$\%$ of the strong-[OI]. 
These results suggest that weak-[OI] LLAGN
have an intermediate age population that is more significant than in
non-active galaxies, and have less contribution of the old population than
in non-active galaxies.

\subsection{Emission line versus stellar population}

In Paper I we found a strong
tendency of the HOBLs to appear preferentially in weak-[OI] LLAGN. As
discussed in \S3.2, this result also applies to the STIS
observations. The detection of these lines implies the presence of a
substantial intermediate age population (e.g. Gonz\'alez Delgado et al
1999). This suggestion is confirmed by the EPS analysis done here,
which shows that LLAGN with conspicuous HOBLs (i.e. those classified
$\eta = $ {\it I}) have $x_8 + x_9 \geq 30\%$. The statistical
connection between these populations and the weak-[OI] class is
clearly illustrated in Figures 10--11, which show that weak-[OI] LLAGN
have sistematically higher contribution of intermediate age populations.
  These results suggest a
connection between LLAGN subtypes and their stellar
population. Because these two classes differ in their emission line
properties, this connection implies a link between the stellar
population and the ionization process in LLAGN.

In Paper I, we have noted a dichotomy between the
[OI]/H$\alpha$ ratio and the spectral indexes, that indicates that
there are no strong-[OI] LLAGN with conspicuous HOBLs and with
$W_K\leq$ 15 \AA\ and $W_C\leq$ 3.5 \AA. These results are also
confirmed by the EPS analysis, and are shown in Figure 12 where we
plot [OI]/H$\alpha$ as a function of the fraction of the light
provided by the old and the intermediate age stellar
populations. The [OI]/H$\alpha$ vs. $x_O$ diagram has an 'inverted L' shape, that can be
interpreted as the strong-[OI] LLAGN having a stellar population
mainly dominated by old stars, while weak-[OI] LLAGN can have stellar
populations of all ages.  However, the [OI]/H$\alpha$ vs. $x_I$  diagram is 'L' shaped, 
showing that many of the weak-[OI] LLAGN have higher $x_I$. 
This also confirms that most of the objects with
intermediate age are weak in [OI]. As we suggested in Paper I, in these
objects  stellar sources can dominate the ionization, and we call them stellar-LLAGN.

We use the STIS spectra to look for these sources. 
The high spatial resolution provided by STIS is crucial to detect
compact young stellar clusters, that otherwise can be masked by the
underlying light emitted by the old stars in ground based
observations.  The EPS analysis finds that only 5 weak-[OI] of the 34 
observed from the ground have a young component with $x_{Y/FC}\geq$
10\%. These objects belong to the {\it I} class; and have $x_I\geq$
30\%. However, the  STIS EPS analysis finds a larger fraction of weak-[OI]
LLAGN with a young stellar component, in particular among those that
belong to the {\it I} class.  In fact, 7 of the 9 weak-[OI] LLAGN in
the {\it I} class have $x_{Y/FC}\geq$ 10\%. In addition, two more weak-[OI]
LLAGN, NGC 5055 and NGC 4459, that belong to the {\it I/O} and {\it O} class
 respectively, have $x_{Y/FC}\geq$ 10\%.  All of
these objects have an important contribution of the intermediate age
stars. It is also important to note that in several objects in which
the young stellar population was not detected in the ground-based
observations, it has been detected in the STIS spectra. This suggests
that some weak-[OI] LLAGN that are classified as {\it I/O} can have a
very compact young cluster that is detected when observed with a small
aperture, as it is the case of NGC 5055.

Further evidence that these objects have young stellar clusters come
 from the UV observations reported by Maoz et al (1998). They found UV
 resonance lines formed in the wind of massive stars in NGC 404, NGC
 4569, and NGC 5055, and possibly in NGC 6500. NGC 3507 also shows 
similar UV features (Gonz\'alez Delgado et al. in preparation).
All these are weak-[OI] LLAGN  and outstand at optical wavelengths for their intermediate age
 population. Note, however, that only NGC 4569 has $x_{Y/FC}\geq$ 10$\%$
in the ground-based aperture. We can thus conclude that
 weak-[OI] LLAGN that belong to the {\it I} or {\it I/O} class, and
 therefore have an important intermediate age population, are
 stellar-LLAGN.

In the strong-[OI] LLAGN, stars cannot play an important role in the gas ionization
because we have not found a noticeable presence of stars younger than
1 Gyr.  Photoionization by an AGN is likely the dominant emission line
mechanism in these sources.  NGC 5005 could be the exception to this,
because it is a strong-[OI] LLAGN with HOBLs, and so has an
intermediate age. 
An AGN as the dominant source of ionization in LLAGN can clearly represented by
the strong-[OI] NGC 3998, NGC 4203, and NGC 4450, for which we have obtained
 $x_O\leq$ 50\%, and $x_{Y/FC}\geq$ 20$\%$.

As pointed out by Cid Fernandes \& Terlevich (1995), a young stellar
population has an optical continuum very similar in shape to a
power-law. There is thus a degeneracy between the young stellar and
the power-law components in the EPS analysis.  This degeneracy can be
broken only at ultraviolet wavelengths.  If the resonance lines of CIV
$\lambda$1550, SiIV $\lambda$1400, NV $\lambda$1240 formed in the wind
of massive stars are detected, then the optical light is provided by a
young starburst. We have retrieved from the HST archive the UV
spectrum of NGC 3998 to search for wind lines, but they are not
present in the spectrum. On the contrary, broad CIV and L$\alpha$
emission lines are detected. Furthermore, the optical spectrum
shows also many FeII multiplets, and the Balmer lines have a broad
component. These spectral features suggest that a Seyfert 1 nucleus
dwells in this strong-[OI] LLAGN. NGC 4203 and NGC 4450 have not been
observed at the UV with the HST, so we can not check for the presence
of wind stellar lines. However, their optical spectra show
double-peaked broad H lines (Ho et al 2000; Shields et al 2000). This
detection is interpreted as the presence of a nuclear accretion disk
in these two galaxies. We thus conclude that the high $x_{Y/FC}$
component derived from the EPS analysis of these two galaxies is due
to a non-thermal optical continuum that can be represented by a
power-law. Therefore, NGC 3998, NGC 4203 and NGC 4450 are strong-[OI]
LLAGN that harbor a dwarf-Seyfert 1 nucleus; they are probably
AGN-LINERs. Note, however, that none of these three galaxies outstands
in the 'inverted L' of Figure 2 in Paper I. As expected, weak AGN can
easily be masked by the contribution of the stellar population in a
typical ground-based aperture observation. So it could be that other
strong-[OI] have less luminous AGNs that those in NGC 3998, NGC 4203
and NGC 4450 contributing to the optical continuum with less
than 10\% of the 4020 \AA\ light in a 0.2$\times$1 arcsec
aperture. NGC 2787 and NGC 4143 are two possible cases 
because they have a broad H$\alpha$ emission component and have been
classified by Ho et al (1997a) as L1.9, but with a less luminous AGN
because $x_{Y/FC}\leq 10\%$.

\begin{figure}
\begin{center}
\end{center}
\caption{[OI]/H$\alpha$ vs fraction of the light provides by: (a) old
(10Gyr), and (b) intermediate (1Gyr+100Myr) age population.  Symbols are as in Figure 9. }
\label{}
\end{figure}

\subsection{Mean age of the stellar population}

In Figure 13 we condense the results of the EPS analysis in the
evolutionary diagram devised by Cid Fernandes \etal (2003b). The trick
consists of grouping the population vector onto just 3 components,
which, because of normalization, are confined to a plane in this 3D
space, such that all components may be visualized in a 2D projection.
We choose to group the $10^6$, $10^7$ yr and power-law components in a
$x_{Y/FC}$ component (as we did in the statistical discussion above)
and plot it against the $10^8$ yr fraction ($x_8$). The two remaining
components, $x_9$ and $x_{10}$, run along a third, perpendicular
axis. The dotted lines in Figure 13  indicate lines of constant $x_{9+10}
\equiv x_9 + x_{10}$, as labeled. With these choices, young systems
are located in the bottom right of this diagram, while $\sim 10^8$ yr
populations appear in the top left and systems dominated by
populations of $10^9$ yr or more occupy the bottom left corner. In
other words, evolution proceeds counter-clockwise in this handy
diagnostic diagram of stellar populations (see Cid Fernandes et al
2003b for details).

A first result which strikes the eye in this plot is the neat
correspondence between the EPS results and the empirical stellar
population classification.  All but one of the $\eta = $ I objects are
located above the $x_{9+10} = 90\%$ line, whereas most of the $\eta =
$ O sources fall bellow this limit, with $\eta = $ I/O objects roughly
in between. This correspondence, previously discussed in our
statitistical considerations (\S4.1), is simply the EPS version of the
excellent relationship between $\eta$ and the input spectral indices.

The distribution of LLAGN in Figure 13 is suggestive of an evolutionary
sequence, similar to that outlined in Paper I to explain
the [OI]/H$\alpha$ versus $W_K$ diagram. In order to
investigate evolutionary effects in more detail we have carried out an
EPS analysis of theoretical galaxy spectra computed with the GISSEL96
evolutionary synthesis code of Bruzual \& Charlot (1993),  processed through our EPS machinery in
exactly the same manner as for the LLAGN data. Details of this
``calibration'' of EPS results in terms of evolutionary synthesis
models are described in Cid Fernandes et al (2003b), where, except for
the inclusion of a power-law in the base, these same calculations were
first discussed.  Models were computed for a pure instantaneous burst
(IB) with ages from 0 to 15 Gyr and for mixtures of an evolving IB and
a fixed 15 Gyr population. These mixed models are parametrized by the 
contrast parameter $c_0$, the fraction of the 4020
\AA\ flux which is associated with the 15 Gyr old population when the
burst starts ($t = 0$). These simplistic scenarios for the
star-formation history provide a useful reference for a qualitative
evaluation of evolutionary effects.

In Figure 14a we overlay these models with the LLAGN data in our
evolutionary diagram. The outer solid line traces the evolution of a
pure IB. The ages (in Myr) of some of the 21 models computed are
indicated.  Other solid lines in this plot correspond to mixed models with contrast
parameters $c_0 = 40$, 60 and 80\% (from left to right,
respectively). Notice that the evolutionary path of mixed models is
similar in shape to that of a pure IB, but with much reduced
amplitudes of the $10^8$ yr and younger components. After $\sim 0.5$
Gyr all models merge in the sequence towards large $x_{9+10}$ in the
bottom left of the plot.

The location of the LLAGN in the above diagram reinforce our
hypothesis that LLAGN line up in what looks to be an evolutionary
sequence, from $\eta = $ I to O. Similar conclusion
can be reached comparing the location of these sources in the
$D_n(4000)$ versus $H\delta_A$ diagram with those of the models of Kauffmann
et al (2003).  The transition from $\eta = I$ to O occurs at 1--2 Gyr,
depending on $c_O$. From the discussion in \S5.1, this is also an
evolutionary sequence in emission line properties, from $\eta = $
I and weak-[OI] to $\eta = $ O and either weak or strong-[OI].

\begin{figure}
\begin{center}
\end{center}
\caption{2D projection of the EPS results. The young plus FC component
is in the horizontal axis and the $10^8$ yr stellar component in the
vertical axis. A third perpendicular axis carries the remaining
fraction of the flux, which goes in stars of $10^9$ yr or more.
Dotted lines represent the loci of constant $x_9+ x_{10}$. Different
symbols represent different $\eta$ classes of LLAGN, as in Figure
9. Arrows mark the position of several LLAGN that have been observed
from the ground (open symbols) and with STIS (filled symbols), and
they show the variation of the stellar population due to the aperture
effect. Ellipses show the uncertainties associated to three representative points:
NGC 4569 (x$_{Y/FC}$= 35$\pm$6 $\%$, x$_{8}$= 41$\pm$7 $\%$);
NGC 5055 (x$_{Y/FC}$= 18$\pm$5 $\%$, x$_{8}$= 14$\pm$6 $\%$); and
NGC 7331 (x$_{Y/FC}$= 3$\pm$1 $\%$, x$_{8}$= 3$\pm$2 $\%$). Error ellipses 
have been computed using the errors and their covariances according to the expressions given by:
http://laeff.inta.es/users/mcs/SED/Theory/slide8.html}
\label{}
\end{figure}

\subsection{Comparison with Seyfert 2s: The role of starbursts in LLAGN}

Our previous investigations of the stellar populations in type 2
Seyferts provide an interesting reference for comparison. Powerful
circumnuclear starbursts have been unambiguously identified in $\sim
40\%$ of nearby Seyfert 2s (Heckman et al 1997; Gonz\'alez Delgado et
al 1998; Storchi-Bergmann et al 1998; Gonz\'alez Delgado, Heckman \&
Leitherer 2001; Joguet et al 2001). These starbursts were originally
detected by means of either UV imaging, UV spectroscopy or the WR
bump. Among other galaxy properties, these starbursts affect the
equivalent width of the CaII K line, which becomes diluted to $W_K <
10$ \AA\ (Cid Fernandes et al 2001a).  Emission line ratios are also
affected. While in ``pure Seyfert 2s'' (defined as nuclei with no
clear signs of starburst activity) line ratios like HeII/H$\beta$ and
[OIII]/H$\beta$ assume a wide range of values, composite systems have
systematically lower values of these ratios. The interpretation is
straightforward.  Since photoionization by a starburst contributes
more to H$\beta$ than to typical Seyfert 2 lines like
HeII$\lambda$4686 and [OIII]$\lambda$5007, the presence of a starburst
around a Seyfert 2 nucleus has the effect of diluting HeII/H$\beta$
and [OIII]/H$\beta$ with respect to the pure AGN values. The resulting
line ratios are in between those of starbursts and ``pure Seyfert
2s''.  In Cid Fernandes et al (2001a) and Storchi-Bergmann et al
(2001) we further speculated that composite starburst + Seyfert 2
systems evolve into ``pure Seyfert 2s''.

The analogies with the results reported in this paper are evident.
Our stellar-LLAGN, i.e. LLAGN with HOBLs, diluted metal absorption
lines and low [OI]/H$\alpha$, qualitatively resemble starburst +
Seyfert 2 composites. Similarly, LLAGN with predominantly old stellar
populations, are analogous to ``pure Seyfert 2s''. In broad terms, the
results reported in this paper sound like a low luminosity extension
of the starburst-AGN connection identified in Seyfert 2s.

These similarities suggest that LLAGN with HOBL harbor a starburst. But do they?
Undoubtedly, the most remarkable result of our survey is that HOBLs
are ubiquitous, if not omnipresent, in weak-[OI] LLAGN. The problem is
that HOBLs are not direct signposts of ongoing
star-formation. Instead, they signal the presence of $10^8$--$10^9$ yr
populations (Gonz\'alez Delgado et al 2001). It is useful to recall
here that for Seyfert 2s we have found that in all cases where young
stars have been conclusively detected by UV data (Heckman et al 1997;
Gonz\'alez Delgado et al 1998), HOBLs are present in the optical. In
other words, every time a young, $< 10^7$ yr starburst is seen in the
UV, a $\ga 10^8$ yr population is identified in the optical by its
deep HOBLs, which points to multiple bursts or $\sim$ continuous
star-formation regime. Inverting and extrapolating from this result,
the ubiquity of HOBLs in weak-[OI] LLAGN could thus be interpreted as
indirect evidence for the existence of young stars, which, as pointed
out above, naturally explains the distribution of points in our
[OI]/H$\alpha$ versus $W_\lambda$ and x$_O$ diagrams.  However, when young
starbursts are present in LLAGN, they are generally faint (see Table 3
and 4). The fact that low luminosity AGN are associated with low
luminosity starbursts may be a trivial consequence of a ``starburst
$\propto$ AGN'' scaling law, which might exist for the mere reason
that both starbursts and super massive black-holes feed on gas. As
pointed out by Ho, Filippenko, \& Sargent (2003), the gaseous content of LINERs and TOs
is small compared to that of Seyferts, which can potentially explain
why starbursts are more easily found in Seyfert 2s than in LLAGN.

Further insight can be gained by comparing the EPS of LLAGN to results
obtained in previous investigations.  In Figure 14b we plot the EPS of
35 Seyfert 2s from Cid Fernandes et al (2001a, renormalized to
$\lambda = 4020$ \AA). Similarly, in Figure 14c we plot the results
for 57 Starburst and HII galaxies from Cid Fernandes et al (2003b).
Filled and open symbols in Figure 14b represent respectively 
starburst + Seyfert 2 composite systems and ``pure Seyfert 2s''.
``Pure Seyfert 2s'' are less luminous than composites, and comprise
both objects dominated by old stars and
sources which present an excess blue continuum whose
nature we cannot assertain with confidence (Storchi-Bergmann et al
2000). The most likely sources for this blue continuum are scattered
light from the AGN and a relatively weak starburst.

Two differences stand out when comparing LLAGN with Seyfert 2s. First,
young starbursts appear far more frequently in Seyfert 2s.  For
instance, $\sim 40\%$ of Seyfert 2s but less than 10\% of the LLAGN
are located aboved the $x_{9+10} = 50\%$ line\footnote{Note, however, 
that we can not exclude here that this difference could be due to a circumnuclear
starburst (100-500 pc) in Seyfert 2 galaxies, 
instead of young stars in the central 100 pc.}. 
In fact, most of the
starburst + Seyfert 2 composites populate the same region as {\it bona
fide} starburst galaxies (Figure 14c). Further evidence that young stars
are more easily detected in Seyfert 2s than in LLAGN is that whereas
the WR bump has been detected in $\sim 10\%$ of the Seyfert 2s in the
samples of Cid Fernandes et al (2001a) and Joguet et al (2001), we
have not detected this feature in any of our LLAGN. This does {\it
not} imply that LLAGN lack young massive stars altogether. After all,
we know from our analysis and UV work (Maoz et al 1998; Colina et al. 2002) 
that at least some LLAGN contain
these stars. Their relative contribution to the optical flux, however,
is clearly much less relevant than in Seyfert 2s. Given the
similarities in the host galaxy properties of Seyferts, LINERs and
TOs, this result implies that starbursts around LLAGN are generally
weaker than in Seyfert 2s.

Secondly, Seyfert 2s are visibly more spread out than LLAGN in our
evolutionary diagram. Interpreted in terms of stellar populations
alone (i.e. neglecting the effects of scattered light), this result
implies that Seyfert 2 have more mixed stellar populations due to
stronger and/or more numerous starbursts in the recent ($\sim 10^8$
yr) past.  The strong clustering of LLAGN along the evolutionary
tracks for ages $\ga 1$ Gyr is simply not seen for Seyfert 2s.

\subsection{Speculations on an evolutionary scenario}

It is further tempting to speculate that at least some Seyfert 2s
evolve into LLAGN.  The intense star-formation in Seyfert 2 composites
must eventually cease with the continuous consumption of the gas
supply. Barring further fuelling events (e.g. interactions), passive
evolution of the stellar populations would move these Seyfert 2s
roughly along evolutionary paths between the pure IB and $c_O = 40\%$
models in Figure 14a, which leads to LLAGN. If star-formation gradually
decays on time-scales of $\sim 10^8$ yr (instead of ceasing abruptly),
after $\sim 10^8$ yr a strong post-starburst (with pronounced HOBLs)
would coexist with an incipient young population. This is the basic
stellar mixture found in the TOs above the $c_O < 40\%$ track in Figure 14a. 
Concomitantly, the UV--blue continuum and emission line
luminosities of the system would decrease steadly because of the drop
in star-formation and accretion rates as gas becomes less and less
available. Luminosities in the red should change less because of the
steady output from the underlying old bulge population. At $\sim 1$
Gyr these systems would reach the ``turn-over'' at $x_{Y/FC} \sim 10$
and $x_8 \sim 30\%$ in Fig 14. HOBLs from $\sim 10^9$ yr populations
would be present. Young stars, if present, should contribute little to
the optical continuum and thus be hard to detect. In another Gyr or so they would
reach the $x_{9+10} \ga 90\%$ zone occupied by $\eta = O$ sources. The
gas excitation in this phase would presumably be dominated by the AGN,
which is consistent with the fact that strong-[OI] sources are
exclusively found in this region. The relative durations of these
phases would become progressively longer along the evolutionary
sequence, which seems to be quantitatively consistent with the
demographics of Seyfert 2s and LLAGN.

One can start this evolution from other initial positions in Figure 14,
although $c_0 \ga 40\%$ is required to turn a composite Seyfert 2 into
$\eta = I$ TOs (compare Figure 14a and b). The ``pure Seyfert 2s''
between $x_{Y/FC} = 10$ and 50\% in Figure 14b could well skip the $\eta
= I$ phase altogether and move directly to $\eta = I/O$ or $O$ LLAGN.
Young stars, if present, would play an irrelevant role in the gas all
the way along this sequence. The change from high to low excitation
emission line-ratios, i.e. from Seyfert 2 to LINER, would be 
attributed to changes in the accretion rate.
In summary, Seyfert 2s contain more younger starbursts than LLAGN.

\begin{figure}
\begin{center}
\end{center}
\caption{Comparison of the EPS results of LLAGN (panel {\it a}),
Seyfert 2s ({\it b}) and Starburst galaxies ({\it c}) in our
``evolutionary diagram''. Dotted lines in all panels mark contours
constant $x_{9+10}$, as labelled  in the left of panel a.  Symbols in
panel a are as in Figure 13.  Solid lines trace the evolution of
an instantaneous burst (IB) in this EPS
diagram. Labels along the outer evolutionary line in panel {\it a}
indicate the age (in Myr). The four different lines correspond to
different constrast ratios $c_O$ of the burst to an underlying 15 Gyr old
population.  The outer line correspond to $c_O = 0\%$, ie, a pure
burst, while the inner lines correspond to $c_0 = 40$, 60 and 80\%
from right to left.  These fractions refer to the time when the burst
starts.}
\label{}
\end{figure}

\section{Summary and Conclusions}

This is the second paper in a series that deals with the study of the 
stellar populations in LLAGN. The goal is to determine
whether the stellar populations in the central 20-100 pc of these galaxies are
related with the ionization processes of the nuclear gas. For this goal, we have
collected STIS optical (2900-5700 \AA) spectra of 28 LLAGN plus 4 HII nuclei
and 1 non-active galaxy from the HST archive. These data are added to the 
ground-based spectra of the LLAGN presented in Paper I. In total, we have 
analyzed a sample of 73 LLAGN, which represents 45\% of this type of objects 
in the HFS97 catalogue. In this second paper we have (1) described the sample;
(2) classified the STIS nuclear (0.2$\times$1 arcsec) stellar population 
by comparing with normal galaxies; (3) measured the equivalent widths
and colors indicative of the stellar populations; (4) performed stellar 
population synthesis (EPS) of the whole sample; (5) correlated EPS results
with the emission line ratio [OI]/H$\alpha$. LLAGN are divided according with 
their [OI]/H$\alpha$ ratio in two subtypes: strong-[OI] ([OI]/H$\alpha\geq$0.25) and weak-[OI] 
([OI]/H$\alpha\leq$0.25). The main results are summarized below:

\begin{enumerate}

\item{Weak-[OI] LLAGN present a higher contribution of the intermediate age population
than strong-[OI]. They also seem to have a younger stellar populationon on average.}

\item{The relation between [OI]/H$\alpha$ and the stellar population suggests a stellar 
origin for the ionization of weak-[OI] LLAGN with HOBL in absorption, although the 
contribution of massive stars is small in the aperture of the ground-based spectra.}

\item{On the other hand, STIS spectra show a larger contribution of young stars in the case of 
weak-[OI] LLAGN than in the ground-based data, suggesting that its reduced contribution is an effect
of the lower contrast of the ground-based data. This result suggests that the dominant source in 
weak-[OI] LLAGN with an intermediate age population is thus stellar. This is confirmed by the 
HST UV spectra available for a few objects. }

\item{On the other hand, the dominant source of ionization in strong-[OI] LLAGN is likely
an AGN, which is consistent with the detection of broad Balmer lines in emission in a few cases 
and a larger contribution of older stars to their stellar population.}

\item{Evolutionary models for the stellar population suggest an age sequence in which the weak-[OI]
LLAGN are younger objects than the strong-[OI]. The evolution proceeds as the small starburst 
in the nuclei of the weak-[OI] fades away decreasing the stellar contribution to the intensity
of H$\alpha$ and leading to a larger [OI]/H$\alpha$ ratio.}

\item{
The evolutionary scenario proposed here is similar to the
one we have previously proposed for Seyfert 2 galaxies 
(Storchi-Bergmann et al. 2001, Cid Fernandes et al. 2001a), in which a 
Seyfert 2 with a circumnuclear starburst (composite Seyfert 2) 
evolves to a "pure" Seyfert 2 as the starburst fades. The weak-[OI] 
LLAGN seem to be lower luminosity counterparts of composite Seyfert 2 galaxies,
while the strong-[OI] LLAGN would be the pure LINERs. The difference between the Seyfert 2 and 
LINERs would be mainly due to the smaller accretion rate or efficiency in LINERs.}

\end{enumerate}

{\bf Acknowledgments}

We thank Grazyna Stasi\'nska, Miguel Cervi\~no, and Luis Colina their helful comments 
from a through reading of the paper, and the referee for her/his suggestions that help to improve the paper.
RGD, and EP acknowledge support by the Spanish Ministry of Science and 
Technology (MCyT) through grants AYA-2001-3939-C03-01 and AYA-2001-2089. RCF and TSB acknowledge
the support from CNPq and CAPES. Some of these data are from observations 
made with the NASA/ESA Hubble Space Telescope, obtained from the data archive at 
the ST-ECF. Additional data presented here have been taken using ALFOSC, which is owned by 
the Instituto de Astrof\'\i sica de Andaluc\'\i a (IAA) and operated
at the Nordic Optical Telescope under agreement between IAA and the NBIfAFG 
of the Astronomical Observatory of Copenhagen. We are grateful to the IAA director
for the allocation of 5.5 nights of the ALFOSC guaranteed time for this proyect.


\end{document}